\documentclass{desyproc}

\begin{document}
%------------------------------------
\title{Search for streaming dark matter axions or other exotica}

%for single authors the superscripts are optional
\author{{\slshape A. Gardikiotis$^1$, V. Anastassopoulos$^{1}$, S. Bertolucci$^{2}$, G. Cantatore$^{3}$, S. Cetin$^{4}$, H. Fischer$^{5}$, W. Funk$^{6}$, D.H.H. Hoffmann$^{7}$$^{,}$$^{8}$, S. Hofmann$^{9}$, M. Karuza$^{10}$, M. Maroudas$^{1}$, Y. Semertzidis$^{11}$, I. Tkachev$^{12}$, K. Zioutas$^{1}$$^{,}$$^{6}$}\\[1ex]
$^1$University of Patras, Patras, Greece\\$^2$INFN, LNF, Bologna, Italy\\$^{3}$University and INFN Trieste, Italy\\$^{4}$Istanbul Bilgi University, Faculty of Engineering and Natural Sciences, Eyup, Istanbul, Turkeyi\\$^{5}$University of Freiburg, Germany\\$^{6}$CERN, Geneva, Switzerland\\$^{7}$Xi'An Jiaoton University, School of Science, Xi'An, China\\$^{8}$National Research Nuclear University “MEPhI“, Moscow, Russian Federation\\$^{9}$Munich, Germany\\$^{10}$Department of Physics, Center for micro, nano sciences and technologies, University of Rijeka, Croatia, and, INFN Trieste, Italy\\$^{11}$Department of physics, KAIST, and Center for Axion and Precision Physics Research, IBS, Daejeon, Republic of Korea\\$^{12}$INR, Moscow, Russia}

% if the proceedings are available online (e.g. at Indico)
% please enter the contribution ID or file_name below for the DOI
%\contribID{32}
\contribID{familyname\_firstname}

% TO THE CONFERENCE EDITORS: 
% please update the following information      
% before sending the template to the authors
\confID{13889}  % if the conference is on Indico uncomment this line
\desyproc{DESY-PROC-2017-XX}
\acronym{Patras 2017} % if you want the Acronym in the page footer uncomment this line
\doi  % if there is an online version we will register DOIs

\maketitle

\begin{abstract}
We suggest a new approach to search for galactic axions or other similar exotica. Streaming dark matter (DM) could have a better discovery potential because of flux enhancement, due to gravitational lensing when the Sun and/or a planet are aligned with a DM stream [1]. Of interest are also axion miniclusters, in particular, if the solar system has trapped one during its formation. Wide-band axion antennae fit this concept, but also the proposed fast narrow band scanning. A network of detectors can provide full time coverage and a large axion mass acceptance. Other DM searches may profit from this proposal.
\end{abstract}

\section{Introduction}

Axions appear in the solution of Peccei and Quinn, to explain the absence of the CP-violation in quantum chromodynamics. Axions are excellent cold dark matter (CDM) candidates in the mass range around (1-100) $\mu$eV. Beyond that range there are also good dark matter candidates, the so-called axions like particles (ALPs). The axion haloscope technique suggested by P. Sikivie [2] is still a widely used method searching for DM axions. Experimental tests for axions predominantly rely on their electromagnetic coupling resulting in their inverse Primakoff resonant conversion into microwave photons inside a strong magnetic field. This technique, along with more recent ideas, allows to search for this elusive particle by slowly scanning the potential axion rest mass range around  $\sim$ 10$^{-5}$ eV/c$^{2}$ by tuning the resonance frequency of the cavity. The quality factor of the cavity Q defines the width of the resonance (1/Q). It is this very narrow resonance response function of the magnetic axion haloscope, which on the one hand optimizes its sensitivity and on the other hand increases the scanning time accordingly. The expected power generated by axion-photon conversions is proportional to the local axion dark matter density $\rho${$_{{\alpha}}$}. For example, the time required, for a cavity inside the CAST magnet, to reach a sensitivity of g$_{{\alpha}{\gamma}{\gamma}}$ = 10$^{-14}$ GeV$^{-1}$ is $\sim$10 days.\\We present here a new approach for terrestrial axion detection that is based on streaming DM axions. A temporal DM flux enhancement can occur due to gravitational lensing effects by the solar system bodies. This novel detection concept (for axions or other similar exotica) requires a ``wide band'' antenna and/or a fast narrow band scanning scheme [1, 3].

\section{The new concept for relic axions detection}

The search for DM, and in particular for relic axions, has been based on the assumed isotropic halo distribution in our galaxy, with a broad velocity distribution around 240 km/s and an average density of $\sim$0.3 GeV/cm$^{3}$. This target choice might have been one of the reasons behind the non-observation of the axion so far.\\Instead, the proposed new detection concept is based on streaming DM axions which propagate near the ecliptic or streams which are temporally aligned with the Sun$\rightarrow$Earth direction or any axion cluster reaching the Earth. The Sun can focus low speed (0.001c-0.3c) incident particles downstream at the position of the Earth [4]. In the ideal case of perfect alignment stream$\rightarrow$Sun$\rightarrow$Earth, the axion flux enhancement of the stream can be very large ($\sim$10$^{6}$) or even more. Planets are also capable of gravitational lensing for slow moving particles. For example, Jupiter can focus particles with speeds around 10$^{-3}$c, which fits the widely assumed detectable dark matter distribution on Earth [5]. It is this temporally axion signal amplification, which (axion) DM searches might utilize. In fact, streaming DM might have a density, which is $\sim$0.3 to 30\% of the local mean DM density ($\sim$0.3 GeV/cm$^{3}$)[6].\\A DM stream propagating along the Sun-Earth direction could surpass the local DM density, and this will give rise to an unexpectedly large DM flux exposure to an axion haloscope. High precision planet orbital data, spacecraft explorations and laser ranging techniques  put an upper limit for the dark matter density at Earth's location of the order 10$^{5}$ GeV/cm$^{3}$ [7]. Also more recent studies on local dark matter density set similar limits [8]. \\Assuming an alignment of a DM stream towards the Earth via the Sun or a planet, lasts only few minutes, this period is equivalent to few years for a conventional detection scheme. In order to utilize sudden axion burst like phenomena the relic axion antenna should be able \\1) to extend the frequency band range to the maximum and \\2) to decrease the scanning time for each frequency to the minimum. \\Even for a tiny DM stream, the large flux enhancement (in particles due to the Sun's gravitational focusing) can result to a much better detection sensitivity. A network of ``wide band'' antennae with fast narrow axion mass scanning mode could be ideal for this novel detection scheme. 

\section{Streaming dark matter axions }

In a wide variety of axion Dark Matter models, a sizable (or even dominant) fraction of axions is confined in a very dense axionic clumps, or miniclusters, with masses M$\sim$10$^{-12}$M\textsubscript{\(\odot\)}. The axion miniclusters originate from specific density perturbations which are a consequence of non-linear axion dynamics around the QCD epoch [9]. There may be $\sim$10$^{24}$ dense axionic clumps (miniclusters) in the galaxy and their concentration on the solar neighbour being $\sim$10$^{10}$ pc$^{-3}$. Typical miniclusters have radius of $\sim$10$^{7}$ km and their axion density is ~$\sim$10$^{8}$ GeV/cm$^{3}$. In fact, in the course of time a fraction of them has been tidally disrupted and forms streams, with an axion density being ``only'' an order of magnitude larger than the average DM distribution. This clumpy structure can lead to observable signal in femto or micro-lensing missions like LSST project or Gaia, but it is also probable to reveal a potential signal in axion haloscopes [10, 11].\\ In case, an axion minicluster is captured by the solar system a direct encounter of the detector with the axion minicluster will enhance the axion density by about 10$^{5}$x the DM average; The Earth's crossing time of such dense axionic clumps is a few days per year [1,3]. The effect of axion miniclusters may be important for direct DM axion searches and deserves a more thorough study.\\In addition, streaming DM can also be present due to other tidal streams like the Sagittarius stream [3]. It is widely accepted that, in cosmic time scale, the Milky Way disrupts the near Sagittarius (Sgr) Dwarf elliptical galaxy during its multiple passages through the galactic disk [12]. The Sgr DM debris could represent a significant halo sub-structure in the galaxy today, thus affecting also the interpretation of DM direct searches [13]. The Sagittarius dark matter debris in some models induces an energy-dependent enhancement of direct search event rates of as much as $\sim$20 - 45\% and is likely to have a non-negligible influence on dark matter detection experiments [14]. Finally, other (un)predictable streams of DM, including caustics [15], may propagate along a gravitationally favourable direction like one from the Sun or Planet-X towards the Earth, which may temporally enhance the local axion flux. An example of potential  interest  is  the  alignment  (within  5.5$^\circ$)  Galactic  Center$\rightarrow$Sun$\rightarrow$Earth, which repeats once annually (18$^{\text{th}}$December).

\section{Conclusions}

Compared to the widely assumed isotropic DM distribution, streaming DM axions or other particles with similar properties, may be finally the better source for their discovery. The assumed isotropic halo of our galaxy, does not take into account substructures of DM in the form of streams or clumps whose particle density can be as much as 10$^{5}$ GeV/cm$^{3}$. Such a density, is not excluded by local bounds based on planetary precision measurements. Therefore, the possibility of large density fluctuations is likely in our solar system. These fluctuations can boost the local axion density up to several orders of magnitude, and so the axion detectability, provided an axion antenna is sensitive to such axion burst like events. Relic axion or other exotic particles streams and/or tidally disrupted axion miniclusters can be gravitationally focused at the position of the Earth by the Sun and/or a solar planet. Even a tiny flux might get temporally strongly enhanced due to gravitational lensing effects, surpassing thus on the long term the isotropic local DM detection sensitivity [3].  It is worth noticing the observed correlation between solar activity and planetary positions, while similar planetary correlation is also observed, for the Earth's atmospheric electron density (TEC) [16]. Such observations support the assumption that DM axions (or other similar exotica) may well constitute a component of slow speed invisible streams, which in general cannot be predicted except the 18$^{\text{th}}$ December alignment (Earth$\rightarrow$Sun$\rightarrow$Galactic Center). And this, because around this period of the year, the Earth's Ionosphere shows an anomalous high electron density [15]. \\In order to utilize any Earth DM encounter, the haloscope antenna should cover ideally the whole calendar year with a quasi-wideband performance, with the fastest possible scanning mode. The reduced sensitivity should be compensated by the burst like impulse of DM flux. CAST-CAPP is preparing a port for such a possibility by using wide band electronics to be sensitive to a wide axion mass range. Since one single axion haloscope cannot fulfil all these requirements, it is obvious that this proposal can be fully realized by a network of haloscopes or other type of detectors, preferentially distributed around the Globe, in order to secure a possible discovery. Therefore, trapped axion miniclusters and streaming DM axions being gravitationally focused by the Sun and/or a Planet, can become instrumental in axion research.

% ****************************************************************************
% BIBLIOGRAPHY AREA
% ****************************************************************************

\begin{footnotesize}

\end{footnotesize}

% ****************************************************************************
% END OF BIBLIOGRAPHY AREA
% ****************************************************************************

\end{document}